\documentclass[showkeys,12pt, preprint,preprintnumbers,nofootinbib, groupedaddress,superscriptaddress,amsmath,amssymb]{revtex4}
\usepackage[dvipdfmx]{graphicx}
\usepackage{dcolumn}
\usepackage{bm}
\usepackage{amssymb}
\usepackage{amsmath}
\usepackage{epsfig}    
\usepackage{color}
\usepackage{slashed}
\usepackage{hhline}

\def\be{\begin{equation}}
\def\ee{\end{equation}}
\newcommand{\bea}{\begin{eqnarray}}
\newcommand{\eea}{\end{eqnarray}}
\newcommand{\nn}{\nonumber}

\numberwithin{equation}{section}

\begin{document}
\title{Generation of radiative neutrino mass in the linear seesaw framework, charged lepton flavor violation and dark matter}
\author{Arindam Das}
\email{arindam@kias.re.kr}
\affiliation{School of Physics, KIAS, Seoul 130-722, Korea}
\affiliation{Department of Physics \& Astronomy, Seoul National University 1 Gwanak-ro, Gwanak-gu, Seoul 08826, Korea}
\affiliation{Korea Neutrino Research Center, Bldg 23-312, Seoul National University, Sillim-dong, Gwanak-gu, Seoul 08826, Korea}

\author{Takaaki Nomura}
\email{nomura@kias.re.kr}
\affiliation{School of Physics, KIAS, Seoul 130-722, Korea}

\author{Hiroshi Okada}
\email{macokada3hiroshi@cts.nthu.edu.tw}
\affiliation{Physics Division, National Center for Theoretical Sciences, Hsinchu, Taiwan 300}

\author{Sourov Roy}
\email{tpsr@iacs.res.in}
\affiliation{Department of Theoretical Physics, Indian Association for the Cultivation of Science, 2A \& 2B Raja S.C.Mullick Road, Jadavpur, Kolkata 700 032, INDIA}

\date{\today}

\preprint{KIAS-P17018}
\begin{abstract}
We investigate a model with local U(1)$_{B-L}$ and discrete $Z_2$ symmetries where two types of weak isospin singlet neutrinos, vector-like charged lepton and exotic scalar fields are introduced. The linear seesaw mechanism is induced at one-loop level through Yukawa interactions associated with the standard model leptons and exotic fields. We also discuss lepton flavor violation and muon anomalous dipole magnetic moment induced by the new Yukawa interaction. In addition, our model has dark matter candidate which is the lightest $Z_2$ odd neutral particle. We calculate the relic density and constraints from direct detection.
\end{abstract}
\maketitle
\section{Introduction}
Different experiments \cite{T2K, MINOS, DCHOOZ, DayaBay, RENO} on neutrino oscillation phenomena \cite{Beringer:1900zz} are consistently giving the firm indications of the existence of the tiny neutrino mass and flavor mixing.
The existence of the neutrino mass allows us to extend the Standard Model(SM) which is an essential window to search for the new physics. The simplest idea to extend the SM with an SM singlet right handed heavy Majorana 
neutrino was introduced in \cite{Seesaw1, Seesaw2, Seesaw3, Seesaw4}. The heavy right handed Majorana neutrinos create a lepton number violating mass term `for the light neutrinos' through a dimension five operator which can naturally explain the tiny neutrino masses. This procedure
is called the seesaw mechanism. The seesaw scale (the mass scale of the heavy Majorana neutrinos) varies from the electroweak scale to the intermediate scale $(\sim10^{15}~\rm{GeV})$ as the neutrino Dirac Yukawa coupling ($Y_{D}$) varies from the scale of electron Yukawa coupling ($Y_{e}\sim 10^{-6}$) up to that
of the top quark ($Y_{t}\sim 1$). If we consider the scale of the seesaw mechanism at the TeV scale or lower, the Dirac Yukawa coupling ($Y_{D}$) becomes very small ($\mathcal{O}(10^{-6})$) to produce appropriate light neutrino masses as suggested by neutrino oscillation experiments and cosmological observations.

Apart from the seesaw mechanism there is another type of mechanism where a small lepton number violating term plays a key role in generating the tiny neutrino mass. Such mechanism is commonly called as the canonical inverse seesaw 
mechanism \cite{InvSeesaw1, InvSeesaw2}. In this scenario unlike the seesaw mechanism, the light neutrino mass is not obtained by the suppression of the heavy neutrino mass. Due to the smallness of the lepton number violating parameter
the heavy right handed neutrinos are pseudo-Dirac in nature. Their Dirac Yukawa couplings with the SM lepton doublets and the SM Higgs doublet could be order one to produce the light neutrino mass. 

There is another type of TeV scale seesaw model which is called the linear seesaw \cite{Wyler:1982dd, Akhmedov:1995ip, Akhmedov:1995vm, Malinsky:2005bi, Gavela:2009cd, Khan:2012zw, Bambhaniya:2014kga, Kashiwase:2015pra}. This is a simple variation of the canonical 
inverse seesaw model. In linear seesaw model we introduce two heavy right handed SM singlet neutrinos with opposite lepton numbers where four right handed SM singlet Majorana heavy neutrinos are used as in canonical inverse seesaw. It has been shown in \cite{Khan:2012zw} 
that from the vacuum metastability bounds the unknown Dirac Yukawa coupling can be constrained. The vacuum stability bounds on the Dirac Yukawa coupling for the canonical type-I frame-work has been studied in 
\cite{Rodejohann:2012px, Chakrabortty:2012np, Bambhaniya:2016rbb}. In our paper we consider the linear seesaw model where we have the $(13)$ and $(31)$ elements of the neutrino mass matrix are nonzero but $(22)$ and $(33)$ elements to be zero;  here the elements of neutrino mass matrix are considered in the basis of $(\nu_L, N_R^C, S_L)$ where $N_R$ and $S_L$ are SM singlet fermions. Whereas in inverse seesaw model $(13)$ and $(31)$ elements are zero, $(33)$ element is nonzero and $(22)$ may or may not be zero. In our model we generate the $(13)$ and $(31)$ elements of the neutrino mass matrix at the one loop level and study various features of this model.

In our model, we apply an extended gauged B$-$L framework with an additional $Z_2$ parity where we also introduce vector-like charged lepton, two types of weak isospin (which is equivalent to $SU(2)_L$) singlet neutrinos, and new scalar fields.
The one loop induced linear seesaw mechanism is realized by Yukawa couplings associated with SM leptons and new fields. 
These Yukawa couplings also induce muon anomalous magnetic dipole moment (muon $g-2$)  where current measurement indicates $\Delta a_\mu = a_\mu^{\rm exp} - a_\mu^{\rm SM} = (28.8 \pm 8.0)\times 10^{-10}$~\cite{Neut3}, and lepton flavor violating (LFV) processes such as $\ell_i \to \ell_j \gamma$ which is taken as constraints \cite{Lindner:2016bgg}.
In addition, the lightest $Z_2$ odd particle is stable which can be a good candidate of dark matter(DM) if it is neutral \cite{Bhattacharya:2016ysw, Patra:2016ofq,Singirala:2017see}.
Then we discuss relic density and constraint from direct detection for our DM candidate. 

The paper is organized as follows. In Section~II, we introduce our model representing particle contents, new interactions and a neutrino mass matrix where we have studied neutrino masses and mixing in the light of neutrino experimental data. In Section~III, we study lepton flavor violation and muon anomalous dipole magnetic moment. In Section~IV we analyze dark matter physics in the model. In Section~V, we give a conclusion.

\section{Model}
In this model we extend the SM with a $U(1)_{B-L}$ gauge group and a discrete $Z_{2}$ parity. 
The relevant part of the particle content has been 
displayed in Tab.~\ref{particle content}.  
The $N_{R_{i}}$ is the heavy right handed Majorana neutrino with three generations to keep the model free from $U(1)_{B-L}$ anomalies. 
The fermion $S_L$ is also a left handed Majorana heavy neutrino which has three generation and is neutral under $U(1)_{B-L}$ gauge group. 
The iso-singlet charged fermion $E$ is vector-like  with odd $Z_2$ parity. 
We also consider that $E$ also has three generation in our model. 
Notice that the lightest $Z_2$ odd particle is stable and can be a good DM candidate if it is electrically neutral.
\begin{table}[ht]
\begin{center}
\begin{tabular}{c|c|c|c|c c}
      &$SU(2)$&  $U(1)$  & $U(1)_{B-L}$& $Z_{2}$  \\
\hline
${L_{L}(\equiv [\nu_L,\ell_L]^T)}$&2&$-\frac{1}{2}$& $-1$ &$+$\\
$e_{R}$&1&$-1$& $-1$ &$+$\\
\hline
\hline
$N_{R_{i=1,2,3}}$&1&0&$-1$ &$+$\\
$S_{L_{j=1,2,3}}$&1&0&0&$+$\\
$E_{{L, R}_{\alpha=1,2,3}}$&1&$-1$&$-1$ &$-$\\
\hline
\hline
$\Phi$&2&$\frac{1}{2}$&0&$+$\\
$\eta$&2&$\frac{1}{2}$&0&$-$\\
$\chi^{-}$&1&$-1$& $-1$ &$-$\\
$\phi$&1&0& $-1$ &$+$\\
\hline
\end{tabular}
\end{center}
\caption{The relevant part of the particle content}
\label{particle content}
\end{table}

We can write the Lagrangian which is relevant for neutrino mass matrix at tree level as follows: 
\bea
\mathcal{L}_{int} \supset y_{\ell} \overline{{L}_{L}} \Phi e_{R} + y_{N} \overline{{L}_{L}} {\tilde \Phi} N_{R} + y_{NS} \overline{N_{R}} S_{L} \phi+ M_{S} \overline{S_{L}^{C}} S_{L}+ h. c.,
\label{Lag1}
\eea
where the first three terms induce the Dirac mass terms after $\Phi$ and $\phi$ getting VEV, and the fourth term with $M_{s}$ is the lepton number violating Majorana mass term.
  We use the SM Higgs field $\Phi$ as 
 \bea
\Phi= \begin{pmatrix}
            \Phi^{+}\\
            \Phi_{0}
          \end{pmatrix},
\Phi^{\ast}= \begin{pmatrix}
            \Phi^{+^{\ast}}\\
            \Phi_{0}^{\ast}
          \end{pmatrix}, 
\tilde{\Phi}=i\sigma_{2} \Phi^{\ast}=\begin{pmatrix}
            \Phi_{0}^{\ast}\\
            -\Phi^{+^{\ast}}
          \end{pmatrix},
        \Phi^{-} =\Phi^{{+}^{\ast}}
 {\rm and~}  
\eta= \begin{pmatrix}
            \eta^{+}\\
            \eta_0 
          \end{pmatrix}
          \label{ortho}                                 
\eea
where neutral components are written by $\Phi_0\equiv (v+h)/\sqrt2$, and $\phi\equiv (v_\phi+\varphi)/\sqrt2$,
$\eta_0 \equiv \frac{\eta^{Re}+i \eta^{Im}}{\sqrt 2}$ and $\tilde\eta\equiv i\sigma_2 \eta^*$.

After the symmetry breaking one can write the neutrino mass matrix in Eq.~\ref{mass1}
\bea
\mathcal{L}_{mass} = \begin{pmatrix}
                                     \overline{\nu^{C}_{L}} & \overline{N_{R}} & \overline{S_{L}^{C}}
                                      \end{pmatrix}
                                      \begin{pmatrix}
                                     0 & m_{D}^{\ast}  &0\\
                                     m_{D}^{\dagger}& 0 & m_{NS}\\
                                     0&m_{NS}^{T}&M_{S} 
                                      \end{pmatrix}
                                       \begin{pmatrix}
                                     \nu_{L} \\
                                      N_{R}^{C} \\
                                       S_{L}
                                      \end{pmatrix},
                                      \label{mass1}
                                      \eea                                      
where Dirac masses can be written by $m_D = y_{N} \frac{v}{\sqrt{2}}$ and $m_{NS} = y_{NS} v_{\phi}$. The B$-$L symmetry forbids the $(22)$ term in the neutrino mass matrix of Eq.~\ref{mass1}.

At this point it must be pointed out that $\phi$ is a B$-$L charged scalar whose vacuum expectation value (VEV) is denoted by $v_{\phi}$. 
The breaking of the electroweak and B$-$L symmetry is induced spontaneously through the potential:
\bea
V_{1}= m_{H}^{2} \Phi^{\dagger}\Phi+ \frac{\lambda_{1}}{2} ( \Phi^{\dagger}\Phi)( \Phi^{\dagger}\Phi)+ m_{\phi}^{2} \phi^{\dagger}\phi+ \frac{\lambda_{2}}{2} ( \phi^{\dagger}\phi) ( \phi^{\dagger}\phi)+ \frac{\lambda_{12}}{2} ( \Phi^{\dagger}\Phi)( \phi^{\dagger}\phi).
\label{pot1}
\eea
After U(1)$_{B-L}$ breaking, we have $Z'$ boson whose mass is given by $v_\phi$. 
In our analysis, we assume $Z'$ boson is sufficiently heavy evading collider constraints. 
Then the mixing between $Z$ and $Z'$ is essentially given in terms of their masses as 
\begin{align}
\tan\theta_{Z-Z'}\approx \frac{m_Z^2}{m_{Z'}^2}\lesssim {\cal O}(10^{-4})
\end{align}
that is negligible tiny, where we take $m_{Z'}=3.5$ TeV. 
 Since $Z'$ does not contribute to neutrino mass and DM physics, we just assume the gauge coupling for $Z'$ is sufficiently small satisfying the current constraint.
Thus we will not discuss phenomenology of $Z'$.
There are other two scalars $\eta$ and $\chi^{-}$ with odd $Z_{2}$ parity, and the potential term containing the $\eta$ and $\chi^{-}$ can be written as
\begin{align}
V_{2} =& m_{\eta} \eta^{\dagger}\eta + \frac{\lambda_{\eta}}{2} (\eta^{\dagger}\eta)(\eta^{\dagger}\eta)+ m_{\chi^{-}} \chi^{-^{\dagger}}\chi^{-} + \frac{\lambda_{\chi}}{2} (\chi^{-^{\dagger}}\chi^{-})(\chi^{-{\dagger}}\chi^{-})+ \frac{\lambda_{\eta\chi^{-}}}{2} ( \eta^{\dagger}\eta)(\chi^{-^{\dagger}}\chi^{-} ) \nonumber \\
& + \lambda_{\Phi\eta}(\Phi^\dag\Phi)(\eta^\dag\eta)+\lambda'_{\Phi\eta}(\Phi^\dag\eta)(\Phi^\dag\eta) 
+ \frac{\lambda''_{\Phi\eta}}2[(\Phi^\dag\eta)^2+{\rm c.c.}] + \lambda_{\eta\phi}(\eta^\dag\eta)(\phi^\dag\phi) \nonumber \\
 &+  \mu (\Phi^{T}. \eta) \chi^{-} \phi,
\label{pot2}
\end{align}
which is invariant under the prescribed gauge group and the $Z_2$ symmetry\footnote{$\mu (\Phi^{T}. \eta) \chi^{-} \phi=\mu (\Phi^{T} \epsilon \eta) \chi^{-} \phi=\mu \begin{pmatrix} \Phi^{+}& \Phi^{0} \end{pmatrix} \begin{pmatrix} 0&1\\ -1& 0\end{pmatrix} \begin{pmatrix} \eta^{+} \eta^{0}\end{pmatrix}(\chi^{-} \phi)= \mu (\Phi^{+} \eta^{0}\chi^{-}\phi-\Phi^{0} \eta^{+} \chi^{-}\phi) \sim -\mu \Phi^{0} \eta^{+} \chi^{-}\phi$. After symmetry breaking, $-\mu \Phi^{0} \eta^{+} \chi^{-}\phi\supset-\mu\frac{v_{\phi}v}{\sqrt{2}}\eta^{+} \chi^{-}$.}. 
Therefore the complete potential of our system will be given as 
\bea
V_{\rm sys} = V_1 + V_2.
\eea

Using seesaw approximation, from Eq.~\ref{mass1} we can write the effective $3 \times 3$ light neutrino mass matrix as
\bea
m_{\nu} = (m_{D}^{\ast}m_{NS}^{-1})M_{s} (m_{D}^{\ast}m_{NS}^{-1})^{T}.
\label{eig1}
\eea 
\begin{figure}[t]
\begin{center}
\includegraphics[clip, width = 0.5 \textwidth]{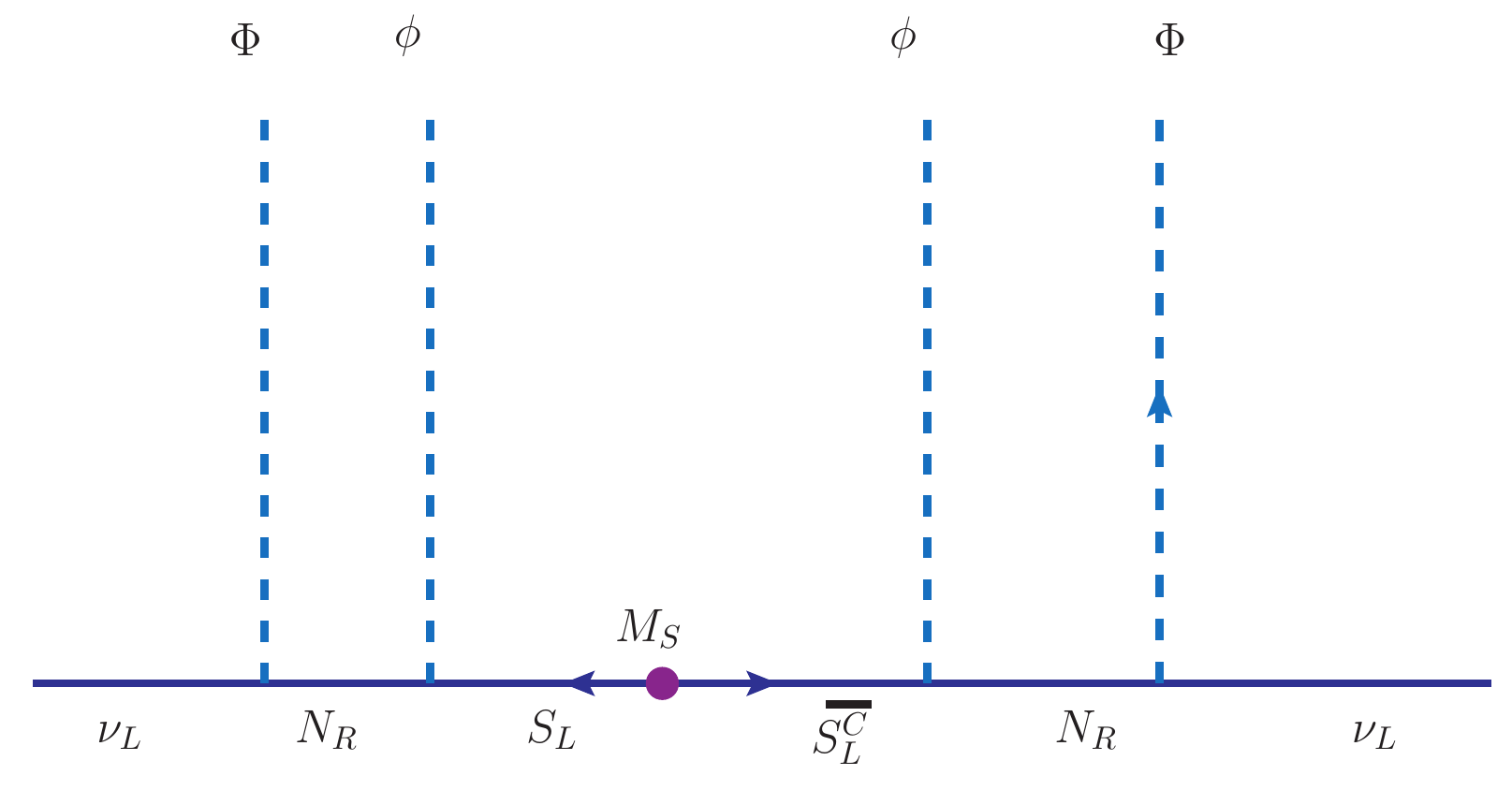}
\caption{Feynman diagram for the inverse seesaw}
\label{ISS}
\end{center}
\end{figure}
                                  
Note that the light neutrino mass is directly proportional to the $M_{s}$. Therefore the degree of smallness regulates the smallness of the light neutrino mass and if $M_{s} \rightarrow 0$, the light neutrino becomes massless, which is the  inverse seesaw scenario \cite{InvSeesaw1, InvSeesaw2}, and the Feynman diagram of the inverse seesaw operator is given in Fig.~\ref{ISS}. 
If there is a non-zero $(22)$ term in \cite{Pilaftsis:1991ug, Dev:2012sg, Dev:2012bd} in the neutrino mass matrix which provides a nontrivial contribution to light neutrino masses at the one loop level which does not vanish in the limit $(33)$ term $(M_S)$ going to zero. However, at the tree level the light neutrino masses go to zero in the limit $M_S \to 0$, even if $M_R \neq 0$.

There is another possibility to obtain the light neutrino mass through switching on the $31$-term in the mass matrix in Eq.\ref{mass1}. This can restore the small neutrino mass even if we have a vanishing $M_{s}$. 
 Here vanishing $M_s$ can be justified by assigning a charge of some global symmetry to $S_L$, $\phi$ and $\chi^-$ as $-1$, $1$ and $-1$, for example,  where only $M_s$ term explicitly breaks the charge conservation in our model. In that case we can interpret that $M_s$ term softly breaks the symmetry and it is natural to take small value for the $M_s$. 
However in our model it is not possible to generate the mass term at the 
tree level because  $U(1)_{B-L}$ symmetry forbids us in writing the terms like $\overline{N_{R}^{C}} N_{R} \phi^{\ast}$, $\overline{{L}_{L}^{c}} \Phi^{\ast} S_{L} $ and $ \overline{e_{R}^{C}} S_L \chi^{+}$ where the first and the second terms respectively induce $22$- and $13(31)$- terms of the neutrino mass matrix while the third term would contribute to a LFV process. Although some terms in neutrino mass matrix are forbidden at tree level, our particle content in Tab.~\ref{particle content} allows 
us to write the Dirac mass term of $E$ and the gauge invariant Yukawa terms which can generate the $13$(or $31$)-term of the neutrino mass matrix through one loop diagram;
\bea
\mathcal{L} \supset  (y_{1})_{i\alpha} \overline{{L}_{L_{i}}} \eta E_{R_{\alpha}} + (y_{2})_{\alpha j} \overline{E_{L_{\alpha}}} S_{L_{j}}^{C} \chi^{-} + M_{E} \overline{E_{L_{\alpha}}} E_{R_{\alpha}},
\label{loop}
\eea 
where $\alpha$ and $j$ is the generation index of the fermions $E$ and $S$ respectively. The third term of Eq.~\ref{loop} is a Dirac mass term of $E$ and will contribute in the neutrino mass generation at one-loop level.           
After generating the $31$(or $13$) term radiatively we can write the neutrino mass matrix~\footnote{It must be mentioned that in Eq.~\ref{mass1} we have three generations of $\nu_L$, three generations of $N_R$ and $S_L$ which makes the Dirac mass matrix, $m_D$ as a $3\times 3$ matrix as $Y_N$ is carrying the flavors. 
The same structure is for Eq.~\ref{mass2} where $\delta_1$ is a $3\times 3$ matrix keeping the other matrices same and the total mass matrix has $9 \times 9$ structure.}
\bea
m_{\nu}^{\rm {tree} + {1-loop}} =                                    
                                      \begin{pmatrix}
                                     0 & m_{D}^{\ast}  & \delta_{1}^{\ast}\\
                                     m_{D}^{\dagger}& 0 & m_{NS}\\
                                      \delta_{1}^{\dagger}&m_{NS}^{T}&M_{S} 
                                      \end{pmatrix}.
                                     \label{mass2}                                      
\eea
\begin{figure}
\begin{center}
\includegraphics[clip, width = 0.5 \textwidth]{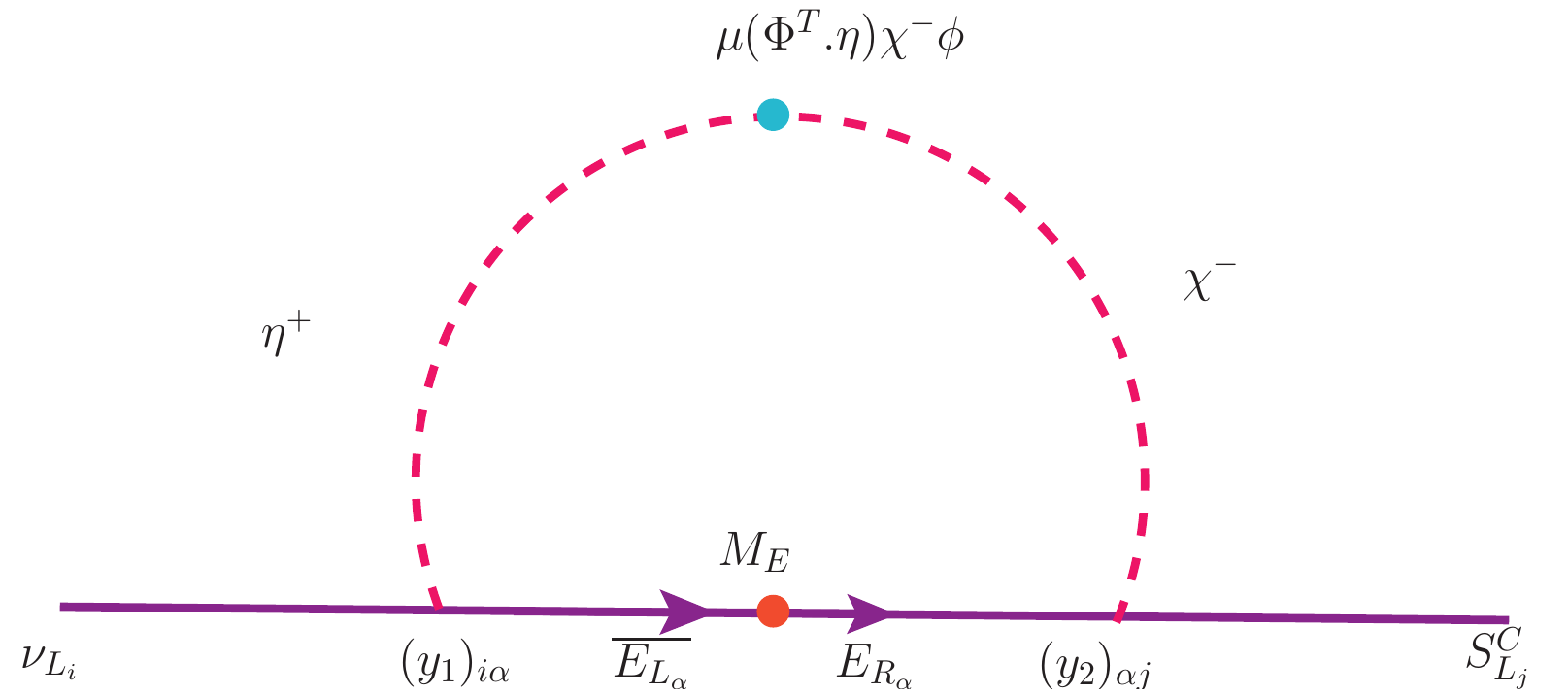}
\caption{Feynman diagram of the radiative loop to generate the $13$(or $31$) mass term in the neutrino mass matrix in Eq.~\ref{mass2} to induce the linear seesaw.}
\label{loop0}
\end{center}
\end{figure}
Using seesaw approximation, from Eq.~\ref{mass2} we can write the effective $3 \times 3$ light neutrino mass matrix as
\bea
(m_{\nu}^{\rm light})^{\rm {tree }+ {1-loop}}&=& \begin{pmatrix}
m_{D}^{\ast}& \delta_{1}^{\ast}
\end{pmatrix}
\begin{pmatrix}
0&m_{NS}\\
m_{NS}^{T}&M_{s}
\end{pmatrix}^{-1}
\begin{pmatrix}
m_{D}^{\dagger}\\
\delta_{1}^{\dagger}
\end{pmatrix} \nonumber \\
&=&\begin{pmatrix}
m_{D}^{\ast}& \delta_{1}^{\ast}
\end{pmatrix}
(\frac{-1}{m_{NS}m_{NS}^{T}})\begin{pmatrix}
M_{s}&-m_{NS}\\
-m_{NS}^{T}&0
\end{pmatrix}
\begin{pmatrix}
m_{D}^{\dagger}\\
\delta_{1}^{\dagger}
\end{pmatrix} \nonumber \\
&=&- (m_{D}^{\ast}m_{NS}^{-1})M_{s} (m_{D}^{\ast}m_{NS}^{-1})^{T}+  \nonumber \\
&& (m_{NS}^{T})^{-1} m_{D}^{\ast} \delta_{1}^{\dagger}+ \delta_{1}^{\ast} m_{D}^{\dagger} (m_{NS}^{-1}).
\label{nu-light}
\eea
Therefore vanishing limit of $M_{s}$ in Eq.~\ref{nu-light} will switch off the tree level mass term and the light neutrino mass term will be generated only from the 1-loop term leading to 
\bea
m_{\nu}^{\rm light^{1-loop}}= (m_{NS}^{T})^{-1} m_{D}^{\ast} \delta_{1}^{\dagger}+ \delta_{1}^{\ast} m_{D}^{\dagger} (m_{NS}^{-1}).
\label{nu-light1}
\eea
Therefore, we can resolve the light neutrino mass through the radiative one loop process in the linear seesaw mechanism.
 The 1-loop diagram in Fig.~\ref{loop0} shows the radiative mass term for the $(13)$ and $(31)$ elements in the neutrino mass matrix.
Now solving the diagram, we can calculate the value of $\delta_{1}$. 
To do this we first rotate the charged scalar sector using an arbitrary orthogonal matrix
\bea
\begin{pmatrix}
\chi^{-} \\
\eta^{-}
\end{pmatrix}                                
=  \begin{pmatrix}
\cos\theta & -\sin\theta  \\                            
\sin\theta&\cos\theta
\end{pmatrix}  
\begin{pmatrix}
H_{1}^{-}\\
H_{2}^{-}
\end{pmatrix}.
\label{ortho}                                 
\eea
From the Fig.~\ref{loop0} and using Eq.~\ref{ortho} we write
\bea
(y_{1})_{i \alpha} \overline{{\nu}_{L_{i}}} \eta^{+} E_{R_{\alpha}} &=& ( y_{1})_{i \alpha} \overline{\nu_{L_i}} E_{R_{\alpha}} (\sin\theta H_{1}^{+}+ \cos\theta H_{2}^{+}), \nonumber \\
(y_{2})_{\alpha j} \overline{E_{L_{\alpha}}} S_{L_{j}}^{C} \chi^{-} &=& (y_{2})_{\alpha j} E_{L_{\alpha}} S_{L_{\alpha}}^{C}(\cos\theta H_{1}^{-}-\sin \theta H_{2}^{-}), 
\eea
\bea
-i(\delta_1)_{ij} &=&\int \frac{d^{4}k}{(2\pi)^{4}} (-iy_{2})_{\alpha j} P_{R} \frac{i(\slashed{k}+ M_{E_{\alpha}})}{(k^{2}-M_{E\alpha}^{2})}(-i y_{1})_{i\alpha} P_{R} \sin\theta \cos\theta \Big(\frac{i}{k^{2}-m_{H_{1}}^{2}}-\frac{i}{k^{2}-m_{H_{2}}^{2}}\Big)\nonumber \\
         &=& \frac{-i \sin\theta \cos\theta}{(4\pi)^{2}} (m_{H_{1}}^{2}-m_{H_{2}}^{2})\sum_{\alpha} (y_{1})_{i\alpha} M_{E_{\alpha}}  (y_{2})_{\alpha j}\nonumber \\ 
         && \int_{0}^{1} dx \int_{0}^{1-x}dy \Big[\frac{1}{x M_{E_{\alpha}}^{2}+ y m_{H_{1}}^{2}+ (1-x-y) m_{H_{2}}^{2}}\Big] \nonumber \\
&=& \frac{-i \sin\theta \cos\theta}{(4\pi)^{2}} (m_{H_{1}}^{2}-m_{H_{2}}^{2})\sum_{\alpha} (y_{1})_{i\alpha} M_{E_{\alpha}}  (y_{2})_{\alpha j}\nonumber\\
    &&\Big[\frac{2\Big(M_{E_{\alpha}}^{2} m_{H_{1}}^{2} \ln[\frac{M_{E_{\alpha}}}{m_{H_{1}}}]+m_{H_{2}}^{2}\Big(m_{H_{1}}^{2}\ln[\frac{m_{H_{1}}}{m_{H_{2}}}]+ M_{E_{\alpha}}^{2}\ln[\frac{m_{H_{2}}}{M_{E_{\alpha}}}]\Big)\Big)}{(M_{E_{\alpha}}-m_{H_{1}})(M_{E_{\alpha}}+m_{H_{1}})(M_{E_{\alpha}}-m_{H_{2}})(M_{E_{\alpha}}+m_{H_{2}})(m_{H_{1}}^{2}-m_{H_{2}}^{2})}\Big]
\label{delta}\nonumber \\
&=&-i\frac{\sin2\theta}{16\pi^2} \sum_{\alpha} (y_{1})_{i\alpha}(y_{2})_{\alpha j} M_{E_{\alpha}}\nonumber \\
&&\Big[\frac{\Big(M_{E_{\alpha}}^{2} m_{H_{1}}^{2} \ln[\frac{M_{E_{\alpha}}}{m_{H_{1}}}]+m_{H_{2}}^{2}\Big(m_{H_{1}}^{2}\ln[\frac{m_{H_{1}}}{m_{H_{2}}}]+ M_{E_{\alpha}}^{2}\ln[\frac{m_{H_{2}}}{M_{E_{\alpha}}}]\Big)\Big)}{(M_{E_{\alpha}}^{2}-m_{H_{1}}^{2})(M_{E_{\alpha}}^{2}-m_{H_{2}}^{2})}\Big],
\eea
where we have assumed 
$M_{E_{\alpha}} \neq m_{H_{1}} \neq m_{H_{2}}$, 
and $m_{H_{1(2)}}$ is defined as the mass of the singly charged boson of $H_{1(2)}^\pm$. 
When we take $M_{E_\alpha} \gg m_{H_{k}}$ ($k=1,2$) typical size of $\delta_1$ is approximately given by
\bea
(\delta_1)_{ij} \sim \frac{\sin 2 \theta}{16 \pi^2} \sum_\alpha \sum_{k}(y_{1})_{i\alpha}(y_{2})_{\alpha j} \frac{m_{H_{i}}}{M_{E_\alpha}} m_{H_{k}},
\label{eq:delta1}
\eea
where $\ln[m_{H_k}/M_{E_\alpha}]$ factor is omitted here.

Depending on the mass scales and the scales of the Yukawa couplings one can justify the degree
of smallness of the mass term $\delta_{1}$ so as to reproduce the light neutrino masses at the correct scale. 

\subsection{Neutrino data}
In this analysis we assume that $(m_{NS}^{T})^{-1}m_{D}^{\ast} << 1$ which allows us to express the flavor eigenstates $\Big(\nu\Big)$ of the light Majorana neutrinos in terms of the 
mass eigenstates of the light $\Big(\nu_m\Big)$ and heavy$\Big(N_m\Big)$ Majorana neutrinos where
\bea
\nu\sim \mathcal{N} \nu_{m}+\mathcal{R} N_{m}.
\eea
For simplicity we may consider $\delta$, $m_{D}$ and $m_{NS}$ are real quantities. Here 
\bea
\mathcal{R}=m_{D}m_{NS}^{-1},~~\mathcal{N} =(1-\frac{1}{2}\epsilon) U_{\rm{PMNS}}, ~~\epsilon &=& \mathcal{R}^{\ast}\mathcal{R}^{T}
\eea
and $U_{PMNS}$ is the usual neutrino mixing matrices which can diagonalize $m_{\nu}$ in the following way
\bea
U_{PMNS}^{T} m_{\nu} U_{PMNS} = diag(m_{1}, m_{2}, m_{3}).
\eea
Due to the presence of $\epsilon$, the mixing matrix $\mathcal{N}$ is non-unitary.  For simplicity we consider that there are three degenerate heavy neutrinos.

We consider a situation where the Dirac mass term carries the flavor,
where as the $\delta$ term is proportional to unity. Therefore
\bea
m_{\nu}= \frac{m_{D}}{m_{NS}}\delta+ \delta \frac{m_{D}}{m_{NS}} =2\delta\frac{m_{D}}{m_{NS}}=2\delta \mathcal{R}= U_{PMNS}^{\ast} D_{\rm{NH/IH}}U_{PMNS}^{\dagger},
\eea
\bea
\mathcal{R}= \frac{1}{2\delta}U_{PMNS}^{\ast} D_{\rm{NH/IH}}U_{PMNS}^{\dagger},
\eea
 where NH(IH) represents the shorthand symbol for ``normal (inverted) hierarchy".
Using the neutrino oscillation data  \cite{Neut5,Neut3}
$\sin^{2}2{\theta_{13}}=0.092$, $\sin^2 2\theta_{12}=0.87$, $\sin^2 2\theta_{23}=1.0$, $\Delta m_{\rm{sol}}^2 = 7.6 \times 10^{-5}$ eV$^2$ and $\Delta m_{\rm{atm}}^2= 2.4 \times 10^{-3}$ eV$^2$
we can write
\bea
D_{NH}&=& 
\begin{pmatrix}
\sqrt{0.1\ast\Delta m_{12}^2}&0&0\\
0&\sqrt{\Delta m_{12}^2}&0\\
0&0&\sqrt{\Delta m_{12}^2+ \Delta m_{23}^2}
\end{pmatrix} \nonumber \\
\label{DNH}
\eea
and
\bea
D_{IH}&=&
\begin{pmatrix}
\sqrt{\Delta m_{23}^2-\Delta m_{12}^2}&0&0\\
0&\sqrt{\Delta m_{23}^2}&0\\
0&0&\sqrt{0.1\ast\Delta m_{23}^2}
\label{DIH}
\end{pmatrix}
\eea
respectively. 
We have expressed $D_{NH}$ in Eq.~\ref{DNH} in terms of $m^2_2 - m^2_1 = \Delta m^2_{12} = 0.9*\Delta m^2_{\rm{sol}}$ and $m^2_2 - m^2_3 = -\Delta m^2_{23} = -\Delta m^2_{\rm{atm}}$
whereas $D_{IH}$ in Eq.\ref{DIH} has been expressed in terms of $m^2_2 - m^2_1 = \Delta m^2_{12} = \Delta m^2_{\rm{sol}}$ and $m^2_2 - m^2_3 = \Delta m^2_{23} = 0.9*\Delta m^2_{\rm{atm}}$.
 Without the loss of generality we can also replace the least eigenvalues by zero for the NH and IH cases, however, the choices of the smallness of these values do not affect the smallness of $\delta_{1}$.

Therefore 
\bea
\mathcal{R}^{\ast}\mathcal{R}^{T}=\frac{1}{4\delta^{2}}U_{PMNS} D_{\rm{NH/IH}} U_{PMNS}^{T} U_{PMNS}^{\ast} D_{\rm{NH/IH}} U_{PMNS}^{\dagger}.
\eea
Using the updated result of the non unitarity matrix from the LFV bounds we can write $\mathcal{N}\mathcal{N}^{\dagger}\sim \bf{1}-\epsilon$.
Due to its non-unitarity, 
 the elements of the mixing matrix ${\cal N}$ 
 are severely constrained by the combined data 
 from the neutrino oscillation experiments, 
 the precision measurements of weak boson decays, 
 and the lepton-flavor-violating decays 
 of charged leptons~\cite{Constraints1, Constraints2, Constraints3, Constraints4, Constraints5}. 
We update the results by using more recent data 
 on the lepton-favor-violating decays~\cite{Adam, Aubert, OLeary}: 
\bea
|{\cal N}{\cal N}^\dagger| =
\begin{pmatrix} 
 0.994\pm0.00625& 1.288 \times 10^{-5} &  8.76356\times 10^{-3}\\
 1.288 \times 10^{-5} & 0.995\pm 0.00625 & 1.046\times 10^{-2}\\
 8.76356 \times 10^{-3}& 1.046 \times 10^{-2} & 0.995\pm 0.00625
\end{pmatrix} .
\eea
Since ${\cal N}{\cal N}^\dagger \simeq {\bf 1} - \epsilon$, 
 we have the constraints on $\epsilon$ such that 
\bea
|\epsilon| =
\begin{pmatrix} 
 0.006\pm0.00625& < 1.288 \times 10^{-5} & < 8.76356 \times 10^{-3}\\
 < 1.288\times 10^{-5} & 0.005\pm 0.00625 & < 1.046 \times 10^{-2}\\
 < 8.76356\times 10^{-3}& < 1.046 \times 10^{-2} & 0.005\pm 0.00625
\end{pmatrix} .
\label{eps}
\eea
The most stringent bound is given by the $(12)$-element 
 which is from the constraint on the lepton-flavor-violating 
 muon decay $\mu \to e \gamma$. Using these bounds we can find the minimum value of 
$\delta_1$ as $\delta_{1_{min}}$$\sim$$\mathcal{O}(10~\rm{eV})$.
\section{Charged Lepton Flavor Violation}
\begin{figure}
\begin{center}
\includegraphics[clip, width = 0.5 \textwidth]{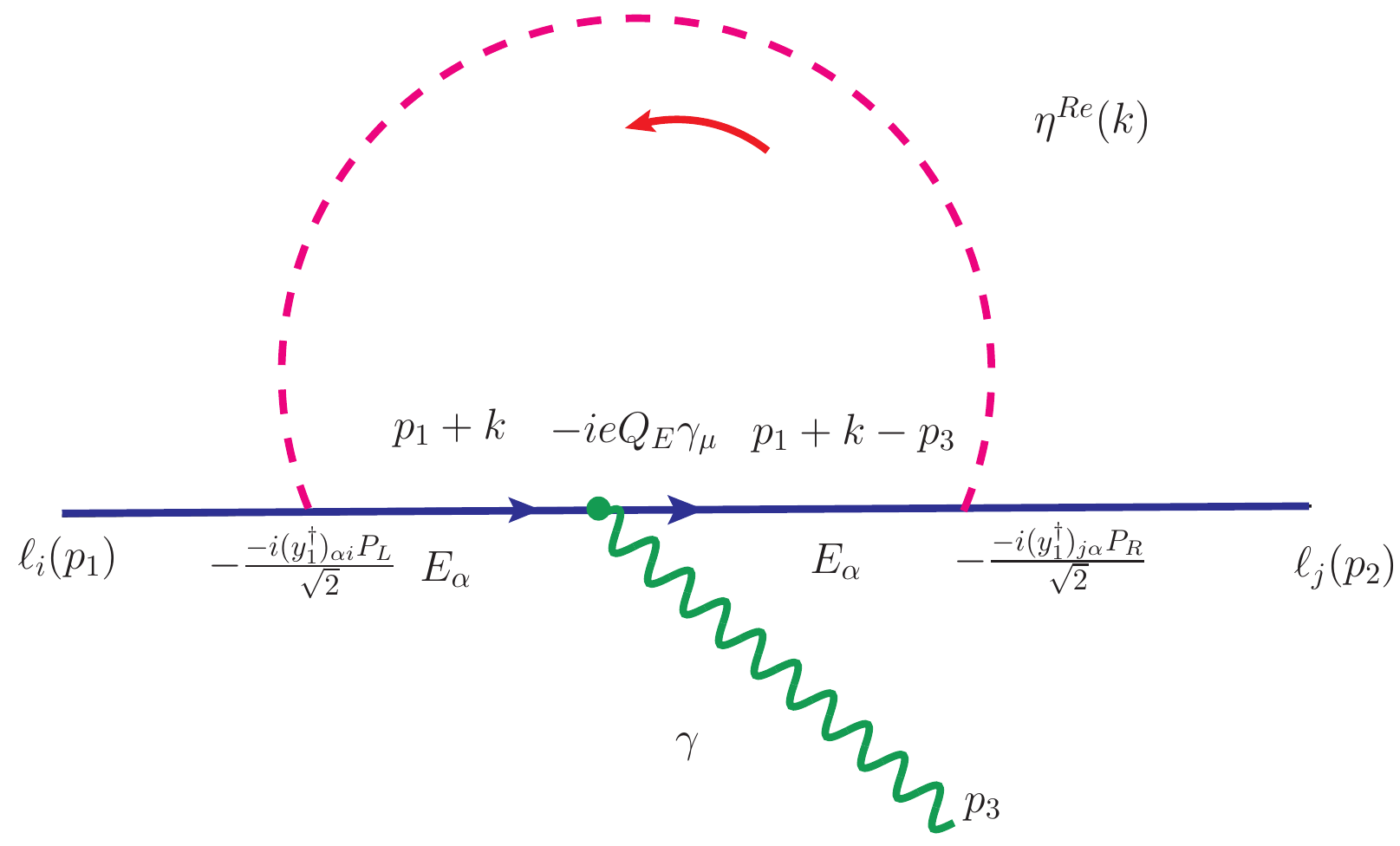}
\caption{Feynman diagram for the charged lepton flavor violation processes $\ell_{i} \to \ell_{j} \gamma$.}
\label{CLFV}
\end{center}
\end{figure}
In our model the fermion $E_{\alpha}$ and the scalar $\eta$ is involved in the charged lepton flavor violation (cLFV)
processes through the interaction 
\bea
\mathcal{L}_{int} \supset \overline{E_{R_\alpha}} (y_{1})^{\dagger}_{\alpha i} {L}_{L_{i}} \tilde{\eta}
 \supset 
\frac{(y_{1_{i\alpha}})^\dag}{\sqrt{2}}  \overline{E_{R_\alpha}} {\ell_{L_i}} (\eta^{{\rm Re}}-i \eta^{{\rm Im}}),\label{eq:clfvs}
\eea
where $\tilde \eta\equiv i\sigma_2 \eta^*$.
The Feynman diagram for the corresponding $\ell_{i} \to \ell_{j} \gamma$ process(es) are given in Fig.~\ref{CLFV}. 
The scattering amplitude for Fig.~\ref{CLFV} is given as\footnote{In our convention $Q_{E}= -1$.} 
{
\bea
i M&=& \int \frac{d^{4}k}{(2\pi)^{4}} \overline{u(p_{2})} \Big[-\frac{-i (y_{1}^{\dagger})_{j \alpha } P_{R}}{\sqrt{2}}\Big] \frac{i(\slashed{k}+\slashed{p_{1}})-\slashed{p_{3}}+{M_{E_\alpha}})}{(k+p_{1}+p_{3})^{2}-{M_{E_\alpha}}^{2}} (i e \gamma_{\mu}) \frac{i(\slashed{k}+\slashed{p_{1}})+{M_{E_\alpha}})}{(k+p_{1})^{2}-{M_{E_\alpha}}^{2}}\nonumber \\
&& \Big[-\frac{-i (y_{1}^{\dagger})_{\alpha i} P_{L}}{\sqrt{2}}\Big] u(p_{1}) \frac{1}{k^{2}-m_{\eta}^{2}} \epsilon(p_{3})^{\mu}\nn \\
&=& i (2 e ~p_{1}.\epsilon^{\ast} ) \overline{u(p_{2})} \Big[a_{R} P_{R}+ a_{L} P_{L}\Big] u(p_{1}),
\eea
where
\bea
&& (a_{R})_{ji} = -{\sum_{\alpha=1}^3 \sum_k^{\rm Re,Im}}
\frac{(y_{1})_{j\alpha}(y^{\dagger}_{1})_{\alpha i}}{2 (4 \pi)^{2}} m_{\ell i} \int dx dy dz\frac{yz\delta(x+y+z-1)}{(x+y){M_{E_\alpha}}^{2}+ z m_{\eta{^k} }^{2}} \\
      &=& -{\sum_{\alpha=1}^3 \sum_k^{\rm Re,Im}}
      \frac{(y_{1})_{j\alpha}(y^{\dagger}_{1})_{\alpha i}}{2 (4 \pi)^{2}} m_{\ell i}\Big[\frac{{M_{E_\alpha}}^6-6{M_{E_\alpha}}^{4} m_{\eta{^k}}^{2}+3 {M_{E_\alpha}}^{2} m_{\eta{^k}}^{4}+2m_{\eta{^k}}^{6}+12 {M_{E_\alpha}}^{2} m_{\eta{^k}}^{4} \ln\Big[\frac{{M}_{\alpha}}{m_{\eta{^k}}}\Big]}
      {12({M_{E_\alpha}^2} -m_{\eta{^k}}^2)^{4}}\Big], \nn
\eea
\bea
&& (a_{L})_{ji} = -{\sum_{\alpha=1}^3 \sum_k^{\rm Re,Im}}
\frac{(y_{1})_{j\alpha}{(y^{\dagger}_1)_{\alpha i}} }{2 (4 \pi)^{2}} m_{\ell j} \int dx dy dz\frac{xz\delta(x+y+z-1)}{(x+y){M_{E_\alpha}}^{2}+ z m_{\eta{^k}}^{2}} \\
      &=&  -{\sum_{\alpha=1}^3 \sum_k^{\rm Re,Im}}
      \frac{(y_{1})_{j\alpha}{(y^{\dagger}_1)_{\alpha i}}}{2 (4 \pi)^{2}} m_{\ell j}\Big[\frac{{M_{E_\alpha}}^{6}-6{M_{E_\alpha}}^{4} m_{\eta{^k}}^{2}+3 {M_{E_\alpha}}^{2} m_{\eta{^k}}^{4}+2m_{\eta{^k}}^{6}+12 {M_{E_\alpha}}^{2} m_{\eta}^{4} \ln\Big[\frac{{M_{E_\alpha}}}{m_{\eta{^k}}}\Big]}{12({M_{E_\alpha}^2} -m_{\eta{^k}}^2)^{4}}\Big]. \nn
\eea
Now
\bea
\mathcal{BR} (\ell_{i}\to \ell_{j}\gamma) \sim \frac{48 \pi^{3} \alpha_{em} C_{ij}}{ G_{F}^{2}m_{\ell i}^{2} } 
{
\left(
\Big|a_{L}^{\eta^{{\rm Re}}}+ a_{L}^{\eta^{{\rm Im}}}\Big|^{2}+ \Big|a_{R}^{\eta^{{\rm Re}}}+a_{R}^{\eta^{{\rm Im}}}\Big|^{2}
\right)_{ji},
}
\eea}
where 
 $\alpha_{em}\approx 1/137$ is the fine structure constant, $G_F\approx1.17\times 10^{-5}$ GeV$^{-2}$ is Fermi constant, and $C_{ij}$
is defined by 
\bea
C_{ij}&\approx&1~ {\rm for}~ (i,j) = (\mu,e) \nonumber \\
&\approx & 0.1784~ {\rm for}~  (i,j) = (\tau,e) \nonumber \\
&\approx & 0.1736~ {\rm for}~(i,j) = (\tau,\mu) .
\eea
The current experimental bound on $\mathcal{BR} (\ell_{i}\to \ell_{j}\gamma)$ is respectively given by~\cite{TheMEG:2016wtm}, \cite{Adam:2013mnn} at 90 \% CL.
\begin{align}
\mathcal{BR} (\ell_{\mu}\to \ell_{e}\gamma)\lesssim 4.2\times 10^{-13},\quad
\mathcal{BR} (\ell_{\tau}\to \ell_{e}\gamma)\lesssim 3.3\times 10^{-8},\quad
\mathcal{BR} (\ell_{\tau}\to \ell_{\mu}\gamma)\lesssim 4.4\times 10^{-8}.
\end{align}
We can avoid the constraints by choosing the Yukawa coupling $y_1$ so that off-diagonal elements of $a_{L(R)}$ are sufficiently small.

 The diagram in Fig.~\ref{CLFV} also contributes to the muon anomalous magnetic moment $\Delta a_\mu$ when $i=j=2$, and it is given by
\bea
\Delta a_{\mu} = -m_{\mu} \Big[a_{L}^{\eta^{{\rm Re}}}+ a_{R}^{\eta^{{\rm Re}}}+ a_{L}^{\eta^{{\rm Im}}}+a_{R}^{\eta^{{\rm Im}}}\Big]_{22},
\eea
including the real and imaginary parts of the neutral  scalar $\eta$. The current experiments~ \cite{bennett, discrepancy1,discrepancy2} report that  its deviation is $(28.8 \pm 8.0)\times 10^{-10}$.
Taking $M_{E_\alpha} > m_{\eta^k}$, we roughly obtain $\Delta a_\mu \sim \sum_\alpha (y_1)_{2\alpha} (y_1^\dagger)_{\alpha 2} (m_\mu/M_{E_\alpha})^2/(96 \pi^2)$. Thus we find that product of Yukawa coupling $\sum_\alpha (y_1)_{2\alpha} (y_1^\dagger)_{\alpha 2}$ should be order one or larger to obtain sizable $\Delta a_\mu$. In addition, exotic particles are preferred not to be too heavy as $\mathcal{O}(1)$ TeV for getting sizable muon $g-2$.

\section{Dark matter scenario}
\begin{figure}
\begin{center}
\includegraphics[clip, width = 0.5 \textwidth]{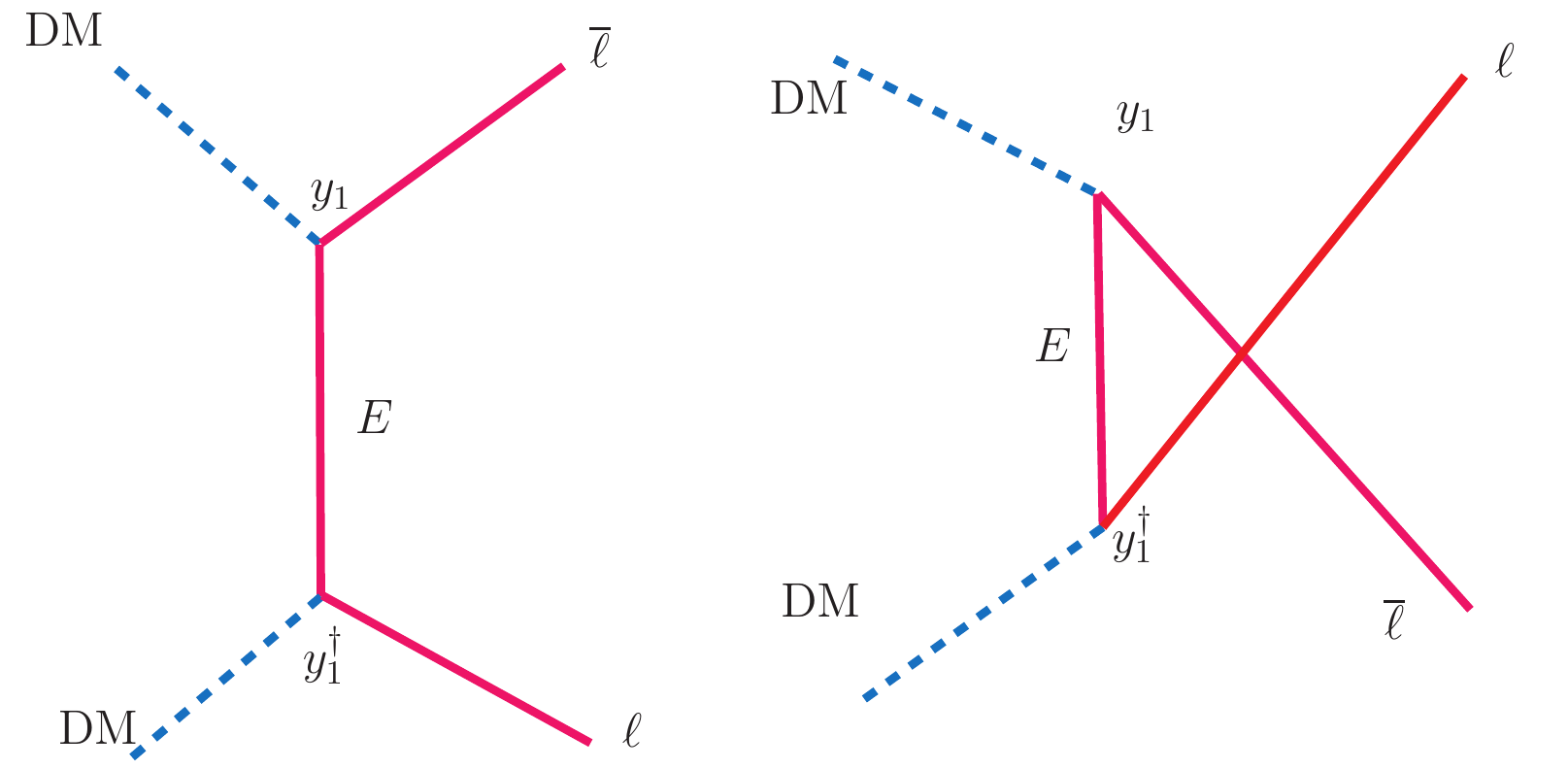}
\caption{Feynman diagram for DM annihilation in $s(t)$-channel (a) and $u$-channel (b).}
\label{DM}
\end{center}
\end{figure}
Neutral component of $\eta$ can be a dark matter (DM) candidate. Here we assume the real part to be DM: $\eta_R\equiv X$.
General analysis has been done by Ref.~\cite{Hambye:2009pw}, where the DM mass is greater than the mass of $W$ boson.~\footnote{In this case, DM mass should be greater than 500 GeV, and coannihilation should also be taken into consideration because of oblique parameter.}
We are interested in lower range $M_X\le m_W$ since it is preferred to obtain sizable muon $g-2$, and thus we focus on this range.  
Also we note that annihilation modes from Higgs portal is subdominant when we require to evade the direct detection constraint such as LUX experiment which is discussed below.
Under this situation, dominant mode comes from the same Yukawa coupling as Eq.\ref{eq:clfvs},
which gives d-wave dominance in the limit of massless final state.
The interaction Lagrangian is again given by
\bea
\mathcal{L}_{int} \supset
  \frac{(y_1)_{i\alpha}}{\sqrt{2}} \overline{\ell_{L_i}} E_{R_\alpha} \Big(\eta^{\rm Re}+ i \eta^{\rm Im}\Big) 
 \supset
   \frac{(y_1)_{i\alpha}}{\sqrt{2}} \overline{\ell_{L_i}} E_{R_\alpha} X .
\eea
The relevant Feynman diagrams for the DM annihilation are given in Fig.~\ref{DM}. Then the nonrelativistic cross section to explain the relic density of DM is obtained by
\begin{align}
\sigma v_{\rm rel}\approx \sum_{i,j = 1}^3 \sum_{\alpha=1}^3\frac{|(y_1)_{i\alpha} (y_1^\dag)_{\alpha j}|^2 M_X^6}{240\pi (M_{E_\alpha}^2+M_X^2)^4} v_{\rm rel}^4 \equiv d_{\rm eff} v_{\rm rel}^4.
\end{align}
Here we apply the relative velocity expansion approximation as follows:
\begin{align}
\Omega h^2\approx \frac{10.7\times 10^{9} \, [{\rm GeV}^{-1}] x_f^3 }{20\sqrt{g_*} M_P d_{\rm eff} },
\end{align}
where $M_P\approx 1.22\times10^{19}$ GeV is the Planck mass, $g_*\approx 100$ is  the total number of effective relativistic degrees of freedom at the time of freeze-out, and $x_f\approx25$ is defined by $M_X/T_f$ at the freeze out temperature ($T_f$), and $d_{\rm eff}$ is the  contribution to the $d$-wave.
 We find that $ \sum |(y_{1})_{i\alpha} (y_1^\dagger)_{\alpha j}|^2$ should be sizable to obtain observed relic density. Note also that even if $y_1$ is large we can obtain small scale of $\delta_1$ in Eq.~\ref{eq:delta1} by small values of $y_2$ and $\theta$.

Spin independent scattering cross section can be found via Higgs portal. The relevant terms in Higgs potential is given in second line of RHS in Eq.~\ref{pot2}.
Then the CP even Higgs mixing in basis of $(\varphi,h)$ is given by
\begin{align}  \begin{pmatrix}
                                     \varphi \\
                                      h
                                      \end{pmatrix}                                
                                =  \begin{pmatrix}
                                     \cos\alpha & -\sin\alpha  \\                            
                                     \sin\alpha&\cos\alpha
                                      \end{pmatrix}  
                                       \begin{pmatrix}
                                      H_1^0\\
                                     H_2^0
                                      \end{pmatrix}  ,
\end{align}%
where $H_2^0$ is the SM Higgs and its mass $m_{H_2^0}\approx125$ GeV, and $H_1^0$ is another neutral Higgs with vacuum expectation value as $v^{\prime}$.
Then its formula is given by
\begin{align}
&\sigma_N \approx \frac{m_N^4}{4(m_N + M_X)^2\pi}\left(\frac{C_{2XH_1^0} s_\alpha}{m_{H_1^0}^2} + \frac{C_{2XH_2} c_\alpha}{m_{H_2^0}^2} \right)^2
\times 3.29\times 10^{-29} {\rm cm^2},\\
&C_{2XH_1^0}=(\lambda_{\Phi\eta} + \lambda'_{\Phi\eta} + \lambda''_{\Phi\eta}) s_\alpha+ \lambda_{\eta\phi} c_\alpha \frac{v_\phi}{v},\nn\\
&C_{2XH_2^0}=(\lambda_{\Phi\eta} + \lambda'_{\Phi\eta} + \lambda''_{\Phi\eta}) c_\alpha- \lambda_{\eta\phi} s_\alpha \frac{v_\phi}{v},\nn
\end{align}
where $m_N\approx0.939$~GeV is the neutron mass.
Here we give a brief estimation, where we simply fix several parameters as $\lambda\equiv \lambda_{\Phi\eta} \approx \lambda_{\Phi\eta} '\approx \lambda_{\Phi\eta}''\approx \lambda_{\eta\phi} $ and $m_{H^0}\equiv m_{H_1^0}\approx m_{H_2^0}=125$ GeV. 
Then the resulting cross section is simplified as
\begin{align}
&\sigma_N \approx \frac{9 \lambda^2 m_N^4}{4\pi(m_N + M_X)^2 m_{H^0}^4}
\times 3.29\times 10^{-29} {\rm cm^2},
\end{align}
notice here that it does not depend on $s_\alpha$, $v$, and $v'$.
The stringent cross section is found to be $\sigma_N\approx2.2\times 10^{-46}$ cm$^2$ at $M_X\approx 50$ GeV reported by LUX experiment~\cite{Akerib:2016vxi}, which are supported by CoGENT~\cite{Aalseth:2014eft} and CREST~\cite{Angloher:2011uu}, although their results are more relaxed.
Therefore in our case, the bound on $\lambda$ is found to be
\begin{align}
\lambda\lesssim  0.022.
\end{align}

Here we discuss order estimation to fit the experimental values such as relic density of DM and muon $g-2$ satisfying LFVs, where notice here that the crucial parameter is $y_1$ and we do not need to include the neutrino sector because of a lot of independent parameters.
First of all, the correct relic density can be achieved by taking $|(y_1)_{i\alpha} (y_1^\dag)_{\alpha j}|$ to be order one, where we expect all the scales of exotic masses are of the order of $100-1000$ GeV. Also sizable muon $g-2$ is achieved if we take $|(y_1)_{21}|^2+|(y_1)_{22}|^2+|(y_1)_{23}|^2$ to be order one. While LFVs restricts some components of $y_1$. For example, the most stringent constraint arises from $\mu\to e\gamma$, and its Yukawa combination $(y_1)_{11}(y_1)_{21}^*+(y_1)_{12}(y_1)_{22}^*+(y_1)_{13}(y_1)_{23}^*$ should be taken to be order ${\cal O}(10^{-4})$ to satisfy this bound, where we respectively take the one-loop function and the mediated fields to be order one and 500 GeV.
Comparing these three combinations, one finds that there are allowed regions by controlling each component of $y_1$.  

{\ Before closing this section we discuss $Z_2$ odd particle production at the LHC. The vector-like charged leptons $E_\alpha$ can be produced via electroweak process $pp \to Z/\gamma \to E^+ E^-$ or $Z'$ exchange in s-channel $pp \to Z' \to E^+ E^-$ where we assume $Z'$ coupling is small and the electroweak process is dominant. Then $E_\alpha$ decays into charged lepton and DM via Yukawa interaction as $E^\pm \to \ell^\pm X$. We thus expect charged lepton plus missing energy signal at the LHC. Thus our $E_\alpha$ production signal is smiler to that of electroweak production of sleptons in supersymmetric models and we can roughly obtain mass limit as $M_E > 500$ GeV from current slepton searches~\cite{ATLAS:2017uun}. Note that the mass limit for our exotic charged scalar boson will be less constrained or similar to that of $E^\pm$; the production cross section of the charged scalar $\eta^\pm$ and $\chi^\pm$ are similar to that of $E^\pm$ while they decay as  $\eta^\pm \to E^{\pm(*)} \nu \to \ell^\pm X \nu$  or $\eta^\pm \to W^\pm X$ and $\chi^\pm \to E^{\pm(*)} S \to \ell^\pm X S$ ($E^{\pm *}$ is off-shell state and depending upon the masses $E^{\pm}$ can be on-shell, too.) which give more particles in final states compared to $E^\pm$ case and the significance of finding charged scalar would be reduced. In Table~\ref{tab:cxE} we summarize the $E_\alpha$ pair production cross section of $pp \to Z/\gamma \to E^+ E^-$ for some benchmark values of $M_E$ which are calculated by {\tt CalcHEP}~\cite{Belyaev:2012qa} with $\sqrt{s}=13$ TeV. Therefore we expect more than 10 events for integrated luminosity 300 fb$^{-1}$ for $M_E \lesssim 1$ TeV. More detailed analysis including simulation study is beyond the scope of this paper and will be done elsewhere.
\begin{table}[t]
\centering {\fontsize{10}{12}
\begin{tabular}{c||cccc}\hline\hline
$M_E$ [GeV] & ~~$750$~~ & ~~$1000$~~ & ~~$1250$~~ & ~~$1500$~~ \\ \hline
$\sigma(pp \to Z/\gamma \to E^+ E^-)$ [fb] & 0.3 & 0.064 & 0.016 & 0.0044 \\ \hline
\end{tabular}%
} 
\caption{The cross sections of $pp \to Z/\gamma \to E^+ E^-$ for some benchmark values of $M_E$. }
\label{tab:cxE}
\end{table}

\section{Conclusion}

In this paper, we have proposed an extension of SM with local U(1)$_{B-L}$ symmetry and discrete $Z_2$ symmetry where exotic leptons and scalar particles are introduced.
In particular, two types of weak isospin singlet neutrinos, $N_{R_{i}}$ and $S_{L_{i}}$ are introduced.

Since $N_{R_{i}}$ is charged under the U(1)$_{B-L}$, it  has to have three generations, due to the anomaly cancelation.
While $S_{L_{i}}$ does not have $B-L$ charge that suggests that the number of flavor for $S_{L}$ can be arbitrary.
 Thus we assume to be three generations of $S_L$ for simplicity.
The model induces linear seesaw mechanism through one loop diagram in which $Z_2$ odd particles propagate, if Majorana mass of $S_{L_{i}}$ is suppressed.   
In addition, the lightest $Z_2$ odd neutral particle can be a good DM candidate.

We have shown a formula for the component of the neutrino mass matrix $\delta_1$ which is generated by one-loop diagram. 
Then the neutrino mass matrix is given by $\delta_1$ and the Dirac mass parameters in our neutrino sector through linear seesaw mechanism. 
To fit the neutrino oscillation data, the order of $\delta_1$ is required to be $\delta_1 \gtrsim \mathcal{O}(10~\rm{eV})$ which can  easily be realized choosing the values of relevant parameters in the formula.
We have also derived formulas of muon $g-2$ and lepton flavor violating decay $\ell \to \ell' \gamma$ at one-loop level.
Furthermore, relic density of DM and DM-nucleon scattering are discussed assuming neutral component of inert doublet scalar is dark matter candidate. 
We then find that our model can accommodate with neutrino oscillation data via linear seesaw mechanism, sizable muon $g-2$, and the relic density of DM, satisfying the constraints from lepton flavor violations and the direct detection experiment of DM.

Such a model can also be tested at the collider. A small value of $\delta_1$ ensures a sizable mixing between the SM light leptons and the BSM fermions. Through such mixings the BSM fermions can be produced at the high energy collider such as Large Hadron Collider (LHC) and 100 TeV pp collider, using $W$ boson and $Z$ boson exchange from the charged current and neutral current interactions respectively. In fact due to the B$-$L model framework, the pair production of such fermions can be tested through the B$-$L gauge boson. These fermions can display the multilepton final states through the corresponding charged current and neutral current interactions \cite{Dev:2013oxa,Dev:2013wba,Das:2014jxa, Das:2015toa, Das:2016akd, Das:2016hof, Das:2017pvt}  which will be interesting in the High Luminosity era of the high energy collider/s. Moreover a general parameter structure can also be adopted for such models as discussed in \cite{Das:2012ze,Das:2017nvm} using the Casas-Ibarra parametrization\cite{Casas:2001sr}. 

In addition, we have several $Z_2$ odd scalars including DM where heavier particles decay into SM leptons and DM via Yukawa interactions. Thus the signals of charged leptons with missing transverse momentum are expected as a signature of these scalar particles. We estimated the cross section of pair production of heavy charged leptons via electroweak process. Then we find that $\mathcal{O}(0.1)$ fb cross section is obtained when heavy charged lepton mass is around 1 TeV. More detailed discussion with simulation is left as future work.

In future a general version of this model under the $U(1)_X$ gauge group can also be considered.  Recently the $U(1)_X$ extended SM has been investigated recently in a variety of contexts, 
such as the classical conformality \cite{Oda:2015gna, Das:2016zue}, $Z^\prime$-portal dark matter \cite{Okada:2016tci}, and 
cosmological inflation scenario \cite{OOR}.

Finally, we also want to comment that such a model can be useful to study baryogenesis via leptogenesis.\cite{Dick:1999je,Murayama:2002je,Abel:2006nv,Bechinger:2009qk,Heeck:2013vha,Ahn:2016hhq, Borah:2016zbd,Blanchet:2010kw, Dev:2014laa, Okada:2012fs} as we can do in the B$-$L, $U(1)_x$, inverse seesaw models. In this model we also have such possibilities to consider three generations of heavy fermions being couples with the SM scalar sector. Such fermions can be non-degenerate, too.Such non-degenerate heavy fermions can have sizable mixings with the SM light neutrinos which are dependent upon the neutrino oscillation data and the free model parameters such as the Dirac phase, Majorana phase, heavy fermion masses and the Casas- Ibarra parametrization. An elaborate discussion on leptogenesis in this model is beyond the scope of this paper and will be considered as a seperate work in the near future.

\bigskip
\acknowledgments
The work AD is supported by the Korea Neutrino Research Center which is established by the National Research Foundation of Korea(NRF) grant funded by the Korea government(MSIP) (No. 2009-0083526).


\end{document}